\documentclass[preprint2]{aastex} 

\shorttitle{Burst Oscillations from NGC 6440}
\shortauthors{Kaaret et al.}

\begin{document}

\title{Discovery of X-Ray Burst Oscillations from a Neutron-Star X-Ray
Transient in the Globular Cluster NGC 6440}

\author{P.\ Kaaret} \affil{Harvard-Smithsonian Center for Astrophysics,
60 Garden St., Cambridge, MA 02138, USA}
\email{pkaaret@cfa.harvard.edu}

\author{J.J.M.\ in 't Zand and J.\ Heise} \affil{SRON National
Institute for Space Research, Sorbonnelaan 2, 3584 CA Utrecht, the
Netherlands \\ Astronomical Institute, Utrecht University, P.O. Box 80
000, 3508 TA Utrecht, the Netherlands}

\author{J.A.\ Tomsick} \affil{Center for Astrophysics and Space
Sciences, Code 0424, 9500 Gilman Drive, \\ University of California at
San Diego, La Jolla, CA 92093}

\begin{abstract}

We report the discovery of millisecond oscillations in an X-ray burst
from the X-ray transient SAX J1748.9-2021 in the globular cluster NGC
6440.   Oscillations at a frequency of $409.7 \pm 0.3$~Hz were present
in one of nine X-ray bursts observed with the Proportional Counter
Array on the Rossi X-ray Timing Explorer during the outburst which
occurred in 2001.  The burst was one of the two dimmest and had the
longest duration and decay time.  The average peak luminosity of two
bursts showing radius expansion is $(3.6 \pm 0.4) \times 10^{38} \rm \,
erg \, s^{-1}$, consistent with the Eddington luminosity for a $1.4
M_{\odot}$ and 10~km radius neutron star burning hydrogen-poor matter.
We speculate that the dichotomy observed between sources with burst
oscillations at once versus twice the frequency difference of kHz
quasiperiodic oscillations in the persistent emission may be related to
the magnetic field geometry of the neutron stars.

\end{abstract}


\section{Introduction}

The discovery of millisecond oscillations in thermonuclear X-ray bursts
provided the first direct measurements of the spin frequencies of
neutron stars in low-mass X-ray binaries (LMXBs)
\citep{strohmayer96,strohmayer02}.  The presence of neutron stars
rotating with millisecond periods in NS-LMXBs was suggested more than a
decade previously \citep{alpar82} and is key in understanding the
relation of LMXBs and millisecond radio pulsars.

X-ray burst oscillation sources form two distinct classes: ``fast
oscillators'' with frequencies close to twice the frequency difference
of the kHz quasiperiodic oscillations (QPOs) seen in the persistent
emission, and ``slow oscillators''  with frequencies near the
difference of the kHz QPO frequencies \citep{white97}.  If the kHz QPO
frequency difference is close to the spin frequency of the neutron
star, then the slow oscillators have a burst oscillation frequency
close to the spin frequency, while the fast oscillators produce signals
at twice the spin frequency. The fast oscillators produce burst
oscillations predominately, but not exclusively, in photospheric radius
expansion bursts, while the slow oscillators produce oscillations in
bursts both with and without photospheric radius expansion
\citep{muno01}.  This indicates that the difference between the two
classes is a physical difference and not a selection effect.  The
nature of this physical difference is unknown.

If the fast oscillators indicate twice the spin frequency and the slow
oscillators indicate the spin frequency, then the spin frequency
distribution of neutron stars in X-ray bursters is clustered with a
maximum value near 400~Hz.  Clustering in the spin frequency
distribution has been variously interpreted as evidence for very
efficient mass ejection by rapidly rotating neutron stars in binaries
\citep{burderi01}, a gravitational radiation limit on the spin-up of
accreting neutron stars \citep{bildsten98} and for the existence of a
phase change in superdense nuclear matter \citep{glendenning01}.

Increasing the sample of sources from which burst oscillations are
detected is crucial for determining the physical nature of the
difference between the fast and slow oscillators and in accurately
measuring the spin frequency distribution of neutron stars in LMXBs.
Here, we describe observations made with the Rossi X-Ray Timing
Explorer (RXTE; Bradt, Rothschild, \& Swank 1993) following the
detection of X-ray bursts from the transient source SAX J1748.9-2021
located in the globular cluster NGC 6440.  We report the discovery of
millisecond oscillations in one X-ray burst.    Because the source is
located in a globular cluster, the distance is known and accurate
luminosities can be calculated.  We find that the peak luminosity of
the radius expansion bursts is consistent with the Eddington limit for
a $1.4 M_{\odot}$ and 10~km radius neutron star with hydrogen-poor
matter. We describe the source and our observations in \S 2, the X-ray
bursts in \S 3, the persistent emission in \S 4, and conclude in \S 5
with a few comments on the nature of the difference between the fast
and slow X-ray burst oscillators.

\section{Observations of NGC 6440}

NGC 6440 is a globular cluster at a distance of $8.5 \pm 0.4$~kpc
\citep{ortolani94} and located near the Galactic center.  Bright,
transient X-ray emission coincident with NGC 6440 was first detected
with OSO-7 and UHURU in December 1971 \citep{markert75,forman76}. 
Strong X-ray outbursts were found, again, in August 1998
\citep{intzand99} and August 2001 \citep{intzand01}.  The peak
luminosities of the outbursts were near $10^{37} \rm \, erg \,
s^{-1}$.  The durations of the outbursts were from one to a few
months.  The 1998 and 2001 outbursts are from the same object
\citep{intzand01} identified with the optical star referred to as V2 in
\citep{verbunt00}.  Whether the 1971 outburst is from the same object
or one of the other, currently quiescent, LMXBs in NGC 6440
\citep{pooley02} is unknown.  Type-I X-ray bursts detected in 1998 and
2001 with the Wide-Field Camera (WFC) on BeppoSAX and a high resolution
image of the cluster obtained with Chandra while the source was active
during 2001 clearly identify the source SAX J1748.9-2021 with V2 and as
an accreting neutron star \citep{intzand01}.  

\begin{figure}[tb] \epsscale{1.0} \plotone{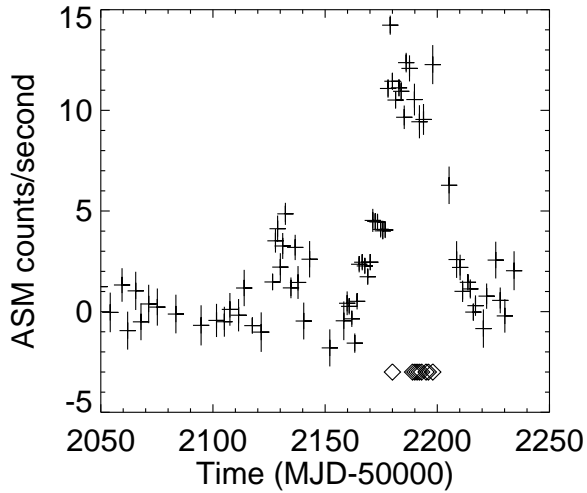} \caption{RXTE/ASM
light curve of the 2001 outburst from NGC 6440.  The points are one day
averages of the ASM count rates.  The diamonds indicate the times of
the PCA observations.} \label{asmlc} \end{figure}

The RXTE All-Sky Monitor (ASM) light curve of the 2001 outburst is
shown in Fig.~\ref{asmlc}.  The outburst started around MJD 52129 (8
August 2001), as reported by the MIT ASM team.  This initial event was
relatively weak at its peak and lasted only 12 days.  After a pause of
about 20 days, NGC 6440 again became bright in x-rays.  The x-ray
emission rose gradually over about 10 days and then jumped rapidly to
the peak flux level near 12 ASM c/s, which corresponds to a flux near
$3.2 \times 10^{-9} \rm \, erg \, cm^{-2} \, s^{-1}$ in the 2--10~keV
band or a luminosity of $3 \times 10^{37} \rm \, erg \, s^{-1}$ for a
source located at the distance of NGC 6440.  High flux levels persisted
for about 20 days, and then the source rapidly declined and became
undetectable with the ASM. 

The X-ray bursts found with the WFC during the 2001 outburst triggered
an RXTE Target-of-Opportunity program which led to an observation on
2001 September 28 and multiple observations during October 7-16.  The
total observation time was 81~ks.   Our observations occurred during
the bright phase of the outburst and in the initial part of the final
decline.  Data were obtained from the Proportional Counter Array (PCA)
in the same spectral and timing modes described in \citet{kaaret02}.

\begin{table*}[tb]
\begin{center}
\begin{tabular}{cccccc}
\tableline
 & Time & Peak Flux & Peak Radius & Decay Time & Duration \\
 &      & $(10^{-8} \rm \, erg \, cm^{-2} \, s^{-1})$ 
                                  & (km)        & (s)        & (s) \\
\tableline
1 & Oct ~8 at 09:20:50 & $5.1 \pm 0.5$ & $48.3 \pm 1.2$ &  3.7 & 11 \\
2 & Oct ~8 at 11:15:05 & $3.8 \pm 0.5$ & $14.3 \pm 0.9$ &  3.9 & 12 \\
3 & Oct 10 at 06:25:36 & $4.6 \pm 0.5$ & $19.9 \pm 1.3$ &  4.3 & 12 \\
4 & Oct 10 at 09:25:03 & $2.5 \pm 0.4$ & $~9.1 \pm 1.1$ &  9.4 & 24 \\
5 & Oct 11 at 07:49:37 & $4.8 \pm 0.5$ & $23.5 \pm 1.5$ &  3.9 & 11 \\
6 & Oct 11 at 09:40:37 & $3.6 \pm 0.5$ & $12.7 \pm 0.9$ &  6.6 & 18 \\
7 & Oct 13 at 07:43:35 & $3.7 \pm 0.5$ & $18.1 \pm 1.4$ &  4.4 & 14 \\
8 & Oct 16 at 04:58:12 & $4.0 \pm 0.6$ & $~9.3 \pm 0.7$ &  6.8 & 21 \\
9 & Oct 16 at 08:40:28 & $2.5 \pm 0.4$ & $~8.5 \pm 0.9$ & 13.0 & 36 \\
\tableline 
\end{tabular}

\caption{Properties of X-Ray Bursts.  The table gives for each burst:
the time (UTC) in the year 2001 at the start the burst, the bolometric
peak flux, the maximum blackbody radius, the $1/e$-folding time in the
tail of the decay, and the duration defined as the interval over which
the flux is greater than 10\% of the peak flux. \label{bursttable}} 
\end{center} \end{table*}

\begin{figure}[tb] \epsscale{1.0} \plotone{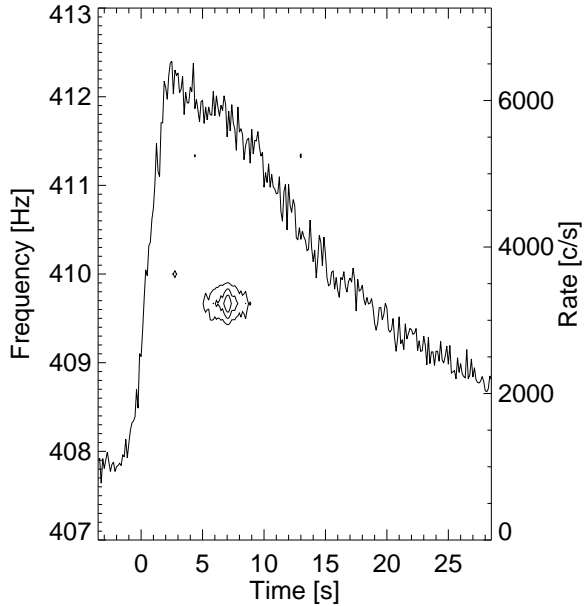} \caption{Dynamical
power spectrum of burst 9 with the burst light curve (count rate)
superimposed.  The contours are at Leahy powers of 12, 20, and 28.  The
contours are generated from power spectra for overlapping 3~s intervals
of data with the power plotted at the mid-point of the interval.  The
left vertical axis gives frequency values for the contours.  The right
axis gives count rates, for the sum of the 3 PCUs on during the burst,
for the light curve.} \label{b9pow} \end{figure}

\section{X-Ray Bursts}

We used the Standard 1 data which has 0.125~s time resolution and no
energy information to search for X-ray bursts and found nine bursts,
see Table~\ref{bursttable}.  We extracted energy spectra for the bursts
from event mode data with 64 channels of energy resolution using all
Proportional Counter Units (PCUs) on during each burst and all layers.
We include data from PCU0 because the problems with background
rejection and modeling caused by the loss of its propane layer are not
important for spectral analysis of these relatively bright bursts.  We
note that we did use the most recent response matrices, which include a
correction for the spectral response due to the loss of the propane
layer.  Spectra were accumulated for 0.25~s intervals and corrected for
deadtime effects.  A spectrum from 10~s of data preceding each burst
was subtracted to eliminate the contribution of the persistent
emission.  

We fitted the resulting spectra in the 3-20~keV band with an absorbed
blackbody model with the column density fixed to $5.9 \times 10^{21}
\rm \, cm^{-2}$, the value equivalent to the optical reddening
\citep{pooley02}.  The results are shown in Table~\ref{bursttable}. 
The bolometric flux was calculated from the spectral fit.  The
equivalent blackbody radius quoted in the table was calculated 
assuming a distance of 8.5~kpc and the uncertainty does not reflect the
distance uncertainty.  We note that these characterizations are only
approximate since significant deviations from blackbody spectra have
been detected from some sources and reprocessing of the radiation in
the neutron star atmosphere alters the observed temperature and
inferred radius from the true values \citep{lewin95}.

Bursts 1 and 5 show evidence for radius expansion including a sharp
increase in radius of more than 20~km and a simultaneous decrease in
temperature of more than 0.7~keV.  The peak fluxes for these two bursts
are consistent within the errors and their average peak luminosity is
$(4.3 \pm 0.5) \times 10^{38} \rm \, erg \, s^{-1}$ where the
uncertainty includes the distance uncertainty.   The fluxes reported
from the RXTE PCA with PCUs 0, 1, 2, and 3 operating (which was the
case during bursts 1 and 5) are systematically a factor of 1.2 higher
than from other instruments.  Assuming that this factor represents an
error in the absolute calibration of the PCA \citep{tomsick99}, we
reduce the flux by this factor and find an average peak luminosity for
these two bursts of $(3.6 \pm 0.4) \times 10^{38} \rm \, erg \, s^{-1}$
which is consistent with the Eddington luminosity for a $1.4 M_{\odot}$
and 10~km radius neutron star with hydrogen-poor matter
\citep{kuulkers02}.  The other bursts have lower peak fluxes.  Burst 3
has a maximum radius close to 20~km and exhibits a temperature drop of
0.8~keV.  Bursts 2, 6 and 7 show weak evidence of radius expansion with
temperature drops of 0.4-0.7~keV but radius expansion of less than
20~km.  Bursts 4, 8 and 9 show no evidence of radius expansion, no
temperature drops, and decay more slowly than the other bursts.

Using high time resolution data (merged event lists from the event and
burst catcher modes) with no energy selection, we computed power
spectra for overlapping 3~s intervals of data, with 0.25~s between the
starts of successive intervals, and searched for excess power in the
range 100-1000~Hz.  We found oscillations in burst 9, with a maximum
Leahy normalized power of 38.7 at a frequency of 409.7~Hz, occurring in
the burst decay about 5~s after the burst peak.  Given 2700 independent
frequencies in the 900~Hz interval searched, the signal power
corresponds to a probability of chance occurrence of $1.1 \times
10^{-5}$, equivalent to a $4.4 \, \sigma$ detection.  The dynamical
power spectrum is shown in Fig.~\ref{b9pow}.  The peak rms amplitude of
the oscillation is near 6\%.  Burst 9 is one of the two dimmest bursts
and has the longest duration and exponential decay time.

\begin{figure}[tb] \epsscale{1.0} \plotone{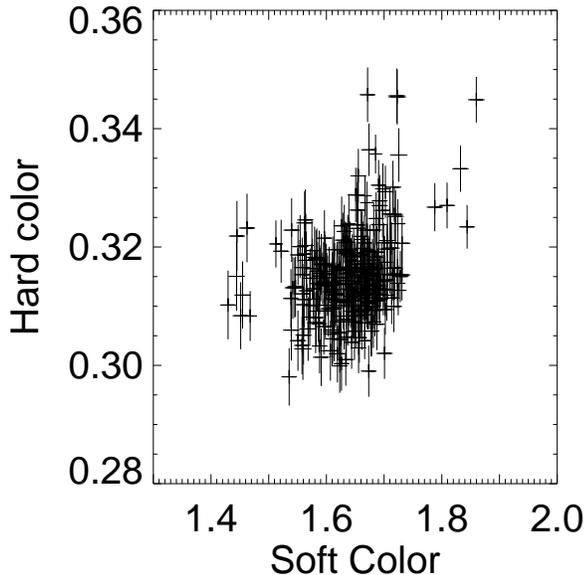} \caption{Color-color
diagram of the persistent emission.  Each point represents 256~s of
data.  The hard color is defined as the count rate in the 9.7--16~keV
band divided  that in the 6.4--9.7~keV band; the soft color is the 
4.0-6.4~keV band rate divided by the 2.6--4.0~keV band rate.}
\label{cc} \end{figure}

\begin{figure}[tb] \epsscale{1.0} \plotone{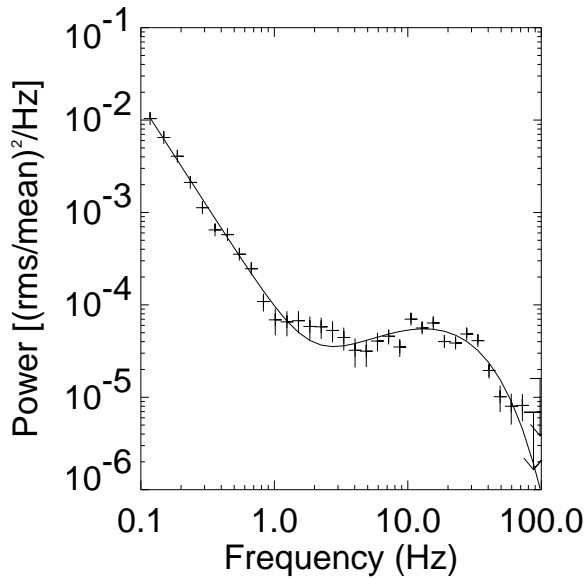} \caption{Low
frequency power spectrum for all observations with soft color above
1.5.  The power is RMS normalized.  The solid line is a fitted curve
described in the text.} \label{lowfreq} \end{figure}

\section{Persistent emission}

We analyzed the persistent emission following the procedures described
in \citet{kaaret02}.  We used only data from PCU2 because this PCU was
on during all of the observations.  PCU0 was also on during all
observations, but a fully reliable background model for PCU0 subsequent
to the loss of its propane layer is not available.  The PCA light curve
shows a high level of variability, up to 80\% within one day.  An X-ray
color-color diagram is shown in Fig.~\ref{cc} with each point
representing a 256~s integration.  The points lie mainly in a single
cluster, except for a few with low soft colors.  A power spectrum for
all of the data with soft color above 1.5 is shown in
Fig.~\ref{lowfreq}.  The power spectrum has very low frequency noise
and a so-called high frequency noise component as typically seen from
atoll sources in the lower banana state \citep{vdk95}.  The solid line
superimposed on the data is a fitted model consisting of the sum of a
low-frequency powerlaw, $\nu^{\gamma}$ with $\gamma = -2.24$, for the
very low frequency noise and an exponentially cutoff powerlaw,
$\nu^{\gamma} e^{-\nu/\nu_{cut}}$ with $\gamma = 0.92$ and $\nu_{cut} =
14.5 \rm \, Hz$.  The RMS fraction is 5.7\% in the 0.1--100~Hz range. 
The points with soft color below 1.5 have the lowest count rates and
are from a single, contiguous time interval.  A power spectrum of these
data  does not appear to have the very low frequency noise component,
which would be consistent with identification as the island state. The
RMS fraction is about 5\% in the 0.1--100~Hz range.

We searched for high frequency QPOs in each uninterrupted RXTE
observation window and in combinations of the various data segments,
including the sum of the island state data.  We calculated averages of
2~s power spectra for all PCA events (2--60~keV) and for events in the
4.7-20.8~keV energy band.  We included events from all PCUs on during
each observation.  We did not find any statistically significant
signals.

\section{Discussion}

Our discovery of burst oscillations from the X-ray transient in NGC
6440 adds another source to the 10 sources previously known to exhibit
burst oscillations.  As described above, X-ray burst oscillation
sources form two classes: ``slow oscillators'' versus ``fast
oscillators'' with the distinction being whether the burst oscillation
frequency is close to once or twice the frequency difference of the kHz
QPOs in the persistent emission \citep{white97}.  The phenomenology of
the fast and slow oscillators is different with the fast oscillators
being much more likely to produce burst oscillations in photospheric
radius expansion bursts which indicates that there is a physical
difference between the two classes \citep{muno01}.  If the kHz QPO
frequency difference is close to the spin frequency of the neutron
star, then the slow oscillators produce signals at close to the spin
frequency and the fast oscillators at twice the spin frequency.


The fast oscillators have frequencies in the range 521-620~Hz, while
the slow oscillators have frequencies in the range 270-363~Hz.  The
410~Hz burst oscillation frequency of SAX J1748.9-2021 lies between the
two classes.  The detection of oscillations during a non-radius
expansion burst and the absence of detections in radius expansion
bursts is more consistent with the phenomenology of the slow
oscillators.  However, conclusive assignment of the source to either
class must wait until kHz QPOs are detected in the persistent emission.
If SAX J1748.9-2021 is a slow oscillator, then it has the highest
frequency of any of the slow oscillators and the highest inferred spin
frequency of any burst source, but it is slower than the spin frequency
of the millisecond X-ray pulsar XTE J1751-305 \citep{markwardt02}.  If
it is a fast oscillator, it has the lowest frequency of any of the fast
oscillators and the lowest inferred spin frequency of any burst source,
but it is faster than the millisecond X-ray pulsar XTE J0929-314
\citep{galloway02}.

The dichotomy between slow and fast oscillators may be related to the
magnetic field configuration of the neutron star.  Diametrically
opposed magnetic poles provide a natural means to produce oscillations
with once or twice the spin frequency.  However, the low upper limits
placed on the subharmonic for fast burst oscillators are a serious
problem for models containing two diametrically opposite hot spots
because of the high degree of symmetry required \citep{muno02}.  The
problem is particularly severe when oscillations are detected early in
the burst rise because nearly simultaneous ignition of the hot spots
would be required.

The evolution of the magnetic field of a neutron star which is
spinning-up or down is determined by interactions between superfluid
neutrons and superconducting  protons within the stellar core
\citep{anderson75}.  For a spinning-up star, quantized magnetic flux
tubes in the superconducting core move inwards towards the rotation
axis \citep{ruderman91}.  For the NS-LMXBs of interest here, it is
likely that sufficient accretion has occurred during spin-up so that
the neutron star crust has been replaced several times, giving a
corresponding movement of the surface magnetic fields even if neither
eddy current dissipation nor creep in the crust has been effective. 
Since accretion is still occurring, the surface magnetic field emerging
from the crust should closely follow the configuration of the core
magnetic field.

For the case where all of the emerging magnetic flux returns to the
same spin hemisphere, the magnetic flux is squeezed into a small polar
cap where the spin axis intersects the crust \citep{chen98}.  For
spin-up from periods near 10~s to periods of milliseconds, both the
north and south magnetic poles move close to one rotational pole and
the net magnetic dipole field becomes nearly orthogonal to the rotation
axis -- an orthogonal rotator.  The local surface magnetic field
strength within the magnetic polar cap is then much higher than the
dipole field, typically $10^{11} - 10^{12} \rm \, G$ for a dipole field
of $10^{8} \rm \, G$.  For the case where flux returns in the opposite
hemisphere, the magnetic field is confined to small polar caps at the
north and south spin poles.  The dipole magnetic field in this spun-up
configuration is oriented nearly parallel to the spin axis -- an
aligned rotator.

Radio millisecond pulsars (MSPs) are similar to NS-LMXBs, indeed they
may be the descendants of the NS-LMXBs, in that they have experienced
significant spin-up subsequent to their formation.  The orientation of
the magnetic field in radio pulsars can be constrained using pulse
profile and polarization measurements.  The fraction of MSPs in the
Galactic disk which are either  orthogonal rotators or nearly aligned
rotators (exactly aligned rotators would not be pulsars, so there must
be a slight offset, $\sim 20\arcdeg$, of the magnetic field from the
rotation axis) is significantly higher than for canonical pulsars
\citep{xilouris98}.  \citet{chen98} showed that this overabundance of
orthogonal and nearly aligned rotators in MSPs can be understood as a
consequence of the magnetic field evolution of a neutron star with a
superfluid and superconducting core, as described above.  The
orthogonal and nearly aligned MSPs represent the two endpoints of the
magnetic field evolution of a spun-up neutron star. 

The magnetic field configuration in LMXBs should be even more tightly
compressed towards the rotation poles than in MSPs because in LMXBs the
neutron star is still actively accreting and spinning up. The same
physical processes which cause the migration of the magnetic poles
toward the rotation axis during spin up cause migration away from the
rotation axis during spin down.  Because MSPs are spinning down, some
motion of the magnetic field away from the poles has occurred (the
magnitude of the motion is small, since the magnitude of the spin down
is small compared to that of the spin up), thereby decreasing the level
of symmetry between the poles.

It is interesting to speculate that the same dichotomy may apply to
NS-LMXBs.  The orthogonal rotator geometry provides a natural means to
produce oscillations at twice the spin frequency, while a slightly
misaligned parallel rotator would produce oscillations at the spin
frequency.  The dichotomy between the fast and slow burst oscillators
would then be related to physical differences between the  magnetic
field configurations of the neutron stars.  

The magnetic field geometry of orthogonal rotators would produce a high
degree of symmetry between the two magnetic poles which would be
located very close to each other.  This would naturally suppress the
signal at the 1/2 subharmonic, i.e.\ the spin frequency, in the fast
oscillators and allow nearly simultaneous ignition since the two poles
are located very close together.  \citet{chen93} estimate that spin up
from 10~s to millisecond periods compresses the magnetic field into a
region with a radius of about $10^{4} \rm \, cm$ around the spin axis. 
For a typical neutron star radius of $10^{6} \rm \, cm$, the poles
would then be less than $1\arcdeg$ from the spin axis and be antipodal
within $2\arcdeg$ as \citet{muno02} conclude is required to explain the
lack of harmonics and sub-harmonics in burst oscillations.  The
requirement that there be no more than 2\% difference in the relative
brightness of the two poles in the main and decaying portions of bursts
\citep{muno02} is more difficult to address, but the high degree of
symmetry between the two poles and their close location for orthogonal
rotators suggest that such uniformity may be achieved several rotation
periods after ignition of the burst.

\acknowledgments  

We greatly appreciate the assistance by the duty scientists of the
BeppoSAX Science Operations Center in the near to real-time WFC data
analysis and the efforts of the RXTE team, particularly Jean Swank and
Evan Smith, in performing these target of opportunity observations.  PK
thanks Mal Ruderman for useful discussions and acknowledges partial
support from NASA grant NAG5-7405.  JZ acknowledges financial support
from the Netherlands Organization for Scientific Research (NWO).



\end{document}